# Measurements of properties of the Higgs-like Particle at 125 GeV by the CMS collaboration

**Sabino MEOLA** *[†]
( *Università di Roma "Guglielmo Marconi" and INFN Napoli* (IT) )
*E-mail*: sabino.meola@cern.ch

CMS results on the measurement of properties of the Higgs-like particle discovered last summer with a mass near 126 GeV are presented. The results are based on a data samples corresponding to an integrated luminosity of up to 5.1 fb$^{-1}$ at 7 TeV and up to 19.6 fb$^{-1}$ at 8 TeV in proton-proton collisions at the LHC. Five decay modes are studied: $\gamma\gamma$, ZZ, WW, $\tau\tau$ and bb. The event yields obtained by different analyses targeting specific decay modes and production mechanisms are consistent with those expected for the standard model (SM) Higgs boson. The mass of the new boson is measured to be 125.7±0.4 GeV. The best-fit signal strength for all channels combined, expressed in units of the SM Higgs boson cross section, is 0.80±0.14 at the measured mass. A discussion on the consistency of the couplings and the spin-parity properties of the observed boson with those predicted for the SM Higgs boson is presented, updated with the most recent results. No significant deviations are found.

---

*Speaker.
[†]A footnote may follow.

## 1. Introduction

Within the Standard Model (SM) of particle physics, the masses of the particles arise from the spontaneous breaking of the electroweak symmetry which is implemented through the Higgs-mechanism. In its minimal version, this is realized through the introduction of a doublet of complex scalar fields. After breaking of the electroweak symmetry, only one scalar field is present in the theory and the corresponding quantum, the Higgs boson, should be experimentally observable. The masses of the fermions in the SM are generated via the Yukawa couplings between the fermions and the Higgs field. Understanding the mechanism for electroweak symmetry breaking is one of the primary goals of the physics program at the Large Hadron Collider (LHC). On July 4th 2012, the CMS experiment (as well as the ATLAS experiment) announced the discovery of a new boson at a mass around 125 GeV, with properties compatible with the SM Higgs boson [1]. A detailed description of the CMS detector is given in [2]. The reported excess is most significant in the SM Higgs searches using the ZZ and $\gamma\gamma$ decay modes. In what follows an updated mass measurement and set of tests on the properties of the new resonance is reported.

## 2. Higgs production and decays

The possible Higgs production modes can be grouped into two main categories: Fermionic production (gluon fusion and top quark pair associated production), and Bosonic production (vector bosons fusion and associated production with a vector boson). The five decays channels are $\gamma\gamma$, ZZ, WW, $\tau\tau$, and bb. The H→$\gamma\gamma$ and H → ZZ → 4l, with l = e,$\mu$, channels play a special role due to the excellent mass resolution of the reconstructed diphoton and four-lepton final states, respectively. The H→WW→$\nu\nu$ channel provides high sensitivity but has relatively poor mass resolution due to the presence of neutrinos in the final state. The bb and $\tau\tau$ decay modes have poor mass resolution and large backgrounds, reducing their sensitivities. In table 1 a summary of expected and measured significance for the 5 channels is shown. All the significances are computed assuming a mass of 125.7 GeV. The bb decay channel still has to be updated with the full statistics. Combining together the $\tau\tau$ and bb channel, the observed significance is 3.4 $\sigma$, which represents the first single experiment evidence of couplings to fermions.

| Decay | Exp. | Obs. |
|---|---|---|
| ZZ | 7.1 $\sigma$ | 6.7 $\sigma$ |
| $\gamma\gamma$ | 3.9 $\sigma$ | 3.2 $\sigma$ |
| WW | 5.3 $\sigma$ | 3.9 $\sigma$ |
| bb | 2.2 $\sigma$ | 2.0 $\sigma$ |
| $\tau\tau$ | 2.6 $\sigma$ | 2.8 $\sigma$ |
| bb+$\tau\tau$ | 3.4 $\sigma$ | 3.4 $\sigma$ |

**Table 1:** Expected and observed significance for the five decay channels and for the bb+$\tau\tau$ channels combined.

## 3. Mass

The combined mass is obtained using the high resolution channels $\gamma\gamma$ and ZZ→4l only. The signal in all channels is assumed to be due to a state with a unique mass and the extraction method is model independent. The description of the general methodology used for the combination can be found in [3, 4]. Results presented are obtained using asymptotic formulae [5], including a few updates recently introduced in the RooStats package [6]. The mass of the new boson is measured to be 125.7±0.3 (stat.) ±0.3 (syst.) GeV. The signal strength modifier $\mu = \sigma/\sigma_{SM}$ represents the measured cross section expressed in unit of the SM Higgs boson cross section.

On the left side of figure 1 is shown the 2-dimensional confidence level (CL) contours for the Higgs boson mass and the signal strength $\mu$, for the high resolution channels $\gamma\gamma$ and 4l and for their combination. The relative signal strengths are constrained by the expectations for the SM Higgs boson. The four main Higgs boson production modes can be associated to either fermion (gg and ttH) or vector boson couplings (VBF and VH). A combination of channels addressing a particular decay mode and explicitly targeting different production modes can be used to test the relative strengths of the couplings to vector bosons and the top quark (right side of figure 1). The most precise result comes from the WW channel, that combines different production modes: associated production, VBF and inclusive. To limit the uncertainty on the x axis, a more accurate measurement in the bb channel is needed.

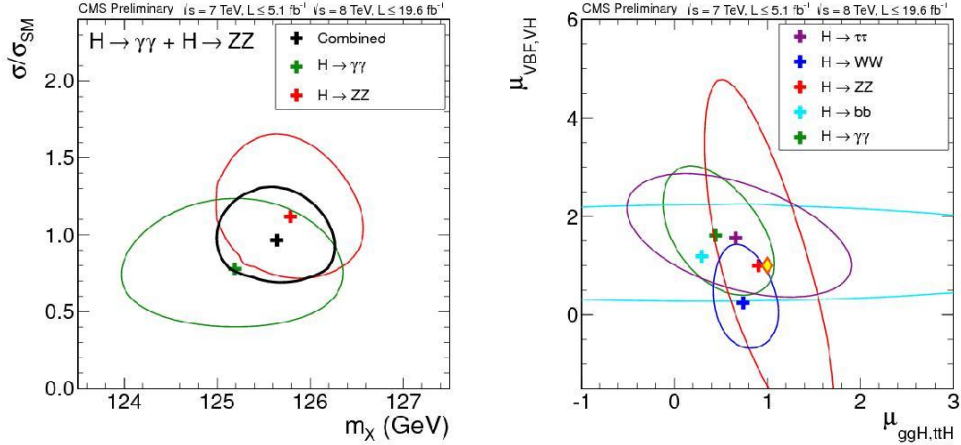

**Figure 1:** Left side: 2D CL contours for Higgs boson mass and signal strength. Right side: fermionic signal strength vs bosonic signal strength.

## 4. Compatibility tests

If the new boson is the SM Higgs boson, all its properties are predicted once its mass is known. To check the consistency with the SM scenario, the total signal strength is computed, which from the combination results equal to 0.8. A CP-even state for the observed resonance is assumed. The expected purities of the different tagged sample vary substantially. If the data are split in different categories enhancing the production mode, a good compatibility is observed with a p-value of 0.5.

Taking the full set of measurements a good compatibility with SM with a p-value of 0.9 is obtained. The ratio of the coupling strengths of the new boson with fermions and vector bosons to the values predicted by the SM is extracted from the various cross section measurements. The formalism to extract the Higgs boson couplings used in this note is the one recommended by the LHC Higgs Cross Section Working Group [7].

The first figure on the left in 2 shows the the CL contours for individual channels and for the overall combination. In the SM the tree level relations between W and Z masses and their couplings to the Higgs boson, are protected against large radiative corrections, a property known as custodial symmetry. Nevertheless in a new physics scenario violations of custodial symmetry are possible. Two scaling factors $k_W$ and $k_Z$ that modify the Higgs coupling to W and Z respectively, are then introduced to assess the consistency of their ratio $\lambda_{WZ}$ with the SM predictions. The ratio of events yield in these channels is model independent because the event are populated essentially by gluon fusion production mechanism. On the central figure in 2, a summary of the results on all the couplings modifier is shown.

A first measurement of spin-parity state of the new particle was reported by CMS [8], disfavoring the pure pseudo-scalar hypothesis over the pure scalar one. Under the assumption of spin 0, data are indeed consistent with the pure scalar hypothesis and disfavor the pure pseudo-scalar hypothesis. The last figure on the right in 2 reports results with improved sensitivity on the $0^+$ vs $2^+_{mgg}$ separation, combining ZZ→4l and WW channels only. Data are consistent with $J^P = 0^+$ within 0.34 $\sigma$ and disfavor $J^P = 2^+$ with a $CL_s$ of 0.6 %.

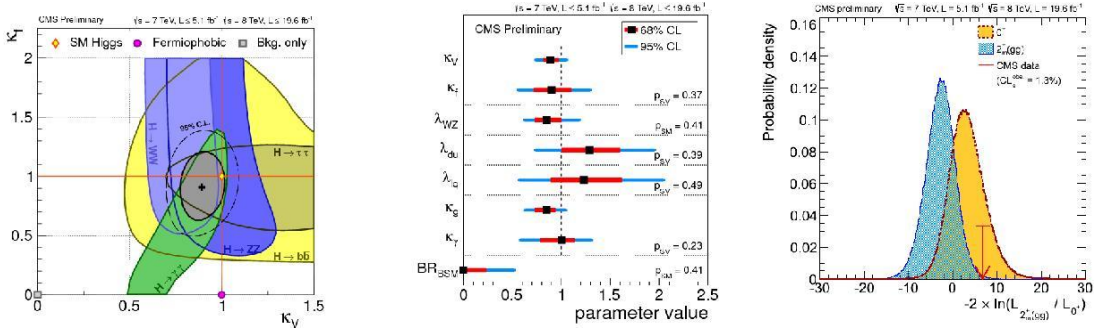

**Figure 2:** From left to right: CL contours for $k_f$ vs $k_v$, summary of coupling modifiers, and $0^+$ vs $2^+_{mgg}$ separation.

## 5. Conclusions

The event yields obtained by the different analyses targeting specific decay modes and production mechanisms are consistent with those expected for the SM Higgs boson. The best-fit signal strength for all channels combined, expressed in units of the SM Higgs boson cross section, is 0.80±0.14 at the measured mass. The consistency of the couplings of the observed resonance with those predicted for the SM Higgs boson is tested in various ways, and no significant deviations are found. Under the assumption that the observed boson has positive parity, the data disfavor the hypothesis of a graviton-like boson with minimal couplings produced in gluon fusion, $J^P = 2^+_m$, with a $CL_s$ value 0.60%. More data is needed to test the SM predicted couplings to better than 20%.